\documentclass[sigconf]{acmart}
\AtBeginDocument{%
  \providecommand\BibTeX{{%
    \normalfont B\kern-0.5em{\scshape i\kern-0.25em b}\kern-0.8em\TeX}}}

\setcopyright{acmcopyright}
\copyrightyear{2021}
\acmYear{2021}

\acmConference[Tran '21]{30th ACM International Conference on Information and Knowledge Management}{November 01--05, 2021}{Gold Coast, Queensland, Australia}

\usepackage{amsmath}
\usepackage[scaled=0.7]{beramono}
\usepackage{makecell}
\usepackage{caption}
\usepackage{subcaption}
\usepackage{tabularx}
\usepackage{multirow}
\usepackage{graphicx}

\DeclareMathOperator{\EX}{\mathbb{E}}% expected value
\begin{document}

\title{ULTRA: An Unbiased Learning To Rank Algorithm Toolbox}

\author{Anh Tran}
\email{abtran@cs.utah.edu}
\affiliation{%
  \institution{University of Utah}
  \streetaddress{ 201 Presidents' Cir}
  \city{Salt Lake City}
  \state{Utah}
  \country{USA}
}
\author{Tao Yang}
\email{taoyang@cs.utah.edu}
\affiliation{%
  \institution{University of Utah}
  \streetaddress{ 201 Presidents' Cir}
  \city{Salt Lake City}
  \state{Utah}
  \country{USA}
}
\author{Qingyao Ai}
\email{aiqy@cs.utah.edu}
\affiliation{%
  \institution{University of Utah}
  \streetaddress{ 201 Presidents' Cir}
  \city{Salt Lake City}
  \state{Utah}
  \country{USA}
}

\renewcommand{\shortauthors}{Tran and Yang, et al.}

\begin{abstract}
Learning to rank system has become an important aspect of our
daily life. However, the implicit user feedback that is used to train
many learning to rank models is usually noisy and suffered from
user bias (i.e., position bias). Thus, obtaining unbiased model using
biased feedback has become an important research field for IR. Existing
studies on unbiased learning to rank (ULTR) can be generalized
into two families-algorithms that attain unbiasness with logged
data, offline learning, and algorithms that achieve unbiasness by
estimating unbiased parameters with real-time user interactions,
namely online learning. While there exist many algorithms from
both families, there lacks a unified way to compare and benchmark
them. As a result, it can be challenging for researchers to choose
the right technique for their problems or for people who are new
to the field to learn and understand existing algorithms. To solve
this problem, we introduced ULTRA, which is a flexible, extensible,
and easily configure ULTR toolbox. Its key features include support
for multiple ULTR algorithms with configurable hyper parameters,
a variety of builtin click models that can be used separately to simulate
clicks, different ranking model architecture and evaluation
metrics, and simple learning to rank pipeline creation. In this paper,
we discuss the general framework of ULTR, briefly describe
the algorithms in ULTRA, detailed the structure, and pipeline of
the toolbox. We experimented on all the algorithms supported by
ULTRA and showed that the toolbox performance is reasonable.
Our toolbox is an important resource for researchers to conduct
experiments on ULTR algorithms with different configurations as
well as testing their own algorithms with the supported features. 
\end{abstract}



\maketitle
\section{Introduction}
As web search engine, e-commerce and multiple different streaming services become an integral part of our daily life, Learning To Rank(LTR) algorithms, the core that powers many of these systems, has grown into an important field of research. One advantage of these algorithms is that they can utilize implicit user feedback such as click data to train models. As a result, researchers in both academia and industry pay great attention to the study of learning to rank using implicit user feedback \cite{10.1145/1076034.1076063}. While user interactions provide us with better understanding of the true utility of each document per user as well as large-scale training data without relying on manual annotations, they usually have inherent noise and bias such as position bias \cite{Ai:2018:ULR:3269206.3274274}. Thus, the LTR research community has created many algorithms to address the problem.

One of the approach to solve the problem is to directly train the model with biased user feedback. This is achieved either by creating learning algorithms that prevent the ranking model for inheriting user bias from observed data or creating interactive online learning process to collect unbiased feedback or estimate unbiased gradient to train the model. This approach is called unbiased learning to rank. Current unbiased learning to rank algorithms can be devised into two categories: offline learning and online learning. Offline learning algorithms focus on preventing models from inheriting observed data, such as search log, bias\cite{10.1145/3209978.3209986,10.1145/3308558.3313447,10.1145/3018661.3018699,10.1145/3159652.3159732}. Online learning algorithms either collect unbiased feedback from real-time user interactions or provide unbiased gradient using the ranking result based on those interactions \cite{Oosterhuis_2018,10.1145/3209978.3210045,10.1145/1553374.1553527}.

While there are many ULTR algorithms from both categories, it lack a unifying and cohesive way to compare and bench mark these algorithms since they are developed using different frameworks, pipelines, logic, models and tested on different datasets. This make choosing the right algorithm for a project very challenging and time consuming. As for researchers who are new to the field, it could be overwhelming and frustrating to have to experiment on various algorithms from different sources.

Our ULTRA toolbox is created to tackle these issues. The algorithms in the toolbox are all developed using Pytorch framework, making them easy to use. In addition, since all the algorithms share the same structure and are implemented using the same pipeline and logic, it is easy to compare them. This also make ULTRA extensible. Adding users' own algorithms or models is quite straightforward as long as they follow the same structure that is currently being implemented in the toolbox. ULTRA is also very flexible. With our API, it is quite intuitive to switch out building blocks such as learning algorithms, models, or click simulator to conduct different experiments, including counterfactual learning algorithms with offline click simulation and online click simulation, learning to rank algorithms with different ranking models and hyper parameters, and online learning algorithms with either deterministic or stochastic online simulation. We experimented on four counterfactual learning algorithms with offline click simulation, deterministic and stochastic online click simulation and four bandit learning algorithms with deterministic and stochastic online click simulation using a multi-layer perceptron network. Out result shows that the ULTRA toolbox produces reasonable results.

The remainder of this paper is organized as follows. Section 2 discuss the existing works related to this paper. Section 3 provides background work and theoretical foundations of existing ULTR algorithms. In section 4, we detail the architecture, classes and pipeline of the toolbox. We provides two use cases that utilize ULTRA to generate unbiased ranking model using Inverse Propensity Weighting algorithm and Duel Bandit Gradient Descent algorithm in section 5. Experiments and results are reported in section 6. In section 7, we discuss how to reproduce our experiment. We summarize and discuss future work in section 8.

\section{Related Work}
Learning to rank is an machine learning approach to train models for ranking tasks. Its application in IR-related areas range from ad-hoc retrieval, Web search, question answering, to recommendation \cite{Ai:2018:ULR:3269206.3274274}. The goal of learning to rank is
to predict a ranking score of an item given its features as inputs. Sorting these scores produces the final ranking list that best capture the system’s or the user’s information need. There are currently two methodologies for ranking models classification based on their structures and definitions. The first one and also the most well known uses models training loss functions as the criteria for categorization. Learning to rank algorithms can be categorized as pointwise, pairwise, or listwise approaches depend on how many items are considered in the loss function for each training step \cite{10.1561/1500000016}. Pointwise approaches essentially take a single document and train a classification or regression model to directly predict the relevance label of the document\cite{inproceedings}. Pairwise approaches look at a pair of documents and try to optimize the ordering of the pair relative to the ground truth\cite{10.1145/1102351.1102363, 10.1145/775047.775067}. Listwise approaches extend pairwise by looking at the entire list of document and try to optimize the ordering of it\cite{10.1145/1273496.1273513,10.1145/3209978.3209985,burges2010from}. The second classification methodology for learning to rank models is to categorize them based on the structure of ranking or scoring functions. Learning to rank algorithms can be classified as univariate method or multivariate method base on the number of items pass into the scoring function in each step. Univariate methods assumption is that documents' relevance are independent of each other. Thus, the learning to rank models' scoring function only need to score one document\cite{10.1561/1500000016}. Multivariate methods assumption, on the other hand, is that documents' context contributes to their ranking. Thus, the scoring functions of learning to rank models take and compare multiple documents together to determine their final ranking scores\cite{10.1145/3209978.3209985,10.1007/978-1-4471-2099-5_24,10.1145/3341981.3344218,10.1145/3397271.3401104,pasumarthi2019selfattentive}.

One of the major challenges is that efficiently train such models often requires large-scale data with annotated relevance labels. These datasets are expensive and time-consuming to create. Thus, IR researchers have tried to employ implicit user feedback as an alternative data source to train ranking model\cite{10.1145/775047.775067}. However, implicit feedback is usually noisy and suffer from different kinds of bias \cite{10.1145/1229179.1229181,O'Brien2006,10.1145/3209978.3210060,10.1145/2484028.2484036}. For example, the position of the document in the ranked list has a strong influence on user decision \cite{10.1145/3130332.3130334}. Documents placed higher in the ranked list tent to have more click. Howerver, this does not neccessary translate to better relevance. Naively train models directly using these implicit feedback could skewer the final ranking result. 

The ULTR approaches aim to produce unbiased models given biased user feedback. As stated above, existing ULTR algorithms that could be categorized into two groups: the counterfactual learning family that originate from offline learning paradigm and the bandit learning family that originate from online learning paradigm. The counterfactual learning's core is Inverse Propensity Weighting(IPW) \cite{10.1145/2911451.2911537} \cite{10.1145/3018661.3018699} and the estimation of examination propensity \cite{10.1145/3209978.3209986,10.1145/3159652.3159732}. Wang
et al. \cite{10.1145/2911451.2911537} propose an experiment that randomized online ranking result to extract the estimated probability that users examine the result of each position and use the estimated weights to debias learning-to-rank models' training loss of. Ai et al. \cite{10.1145/3209978.3209986} build a Duel Learning Algorithm that trains both the ranking model and the examination propensity estimation model together on offline data. Wang et al. \cite{10.1145/3159652.3159732} propose a regression based Expectation-Maximization (EM) algorithm, for personal search, that handle highly sparse click data. There are also algorithms that use online interleaving to estimate propensity \cite{10.1145/3018661.3018699} or collect intervention data from multiple ranking functions \cite{10.1145/3289600.3291017}. There are also counterfactual learning algorithms that implement IPW to different types of behavioral bias such as trust bias \cite{10.1145/3308558.3313697} or recency bias \cite{10.1145/3357384.3358131}.

Bandit learning algorithms take a different approach to create unbias models. The key of bandit learning lies in its ability to estimate unbiased gradients from user feedback as well as online result. One of the popular algorithms is the Dual Bandit Gradient Descent(DBGD) model proposed by Yue and Joachims\cite{10.1145/1553374.1553527}. Its optimization strategy involves updating ranking models' parameters with randomized perturbations iteratively to result in higher user interaction ranking lists. Schuth et al. \cite{10.1145/2661829.2661952} extends DBGD by randomizing multiple perturbed parameters simultaneously in order to speed up convergence. Wang et al. \cite{10.1145/3209978.3210045} propose to only explore null space of recent poorly perform gradients to find a more efficient directions in future steps. Oosterhuis and de Rijke \cite{Oosterhuis_2018} break away from extensive interleaving and multileaving approaches through ranking list sampling and weights updating during training.

The above-mentioned algorithms are independent from each other and can be used to achieve unbiased ranking models. Our toolbox allows the users to choose and configure different algorithms, models and evaluation metrics to fit their needs and provides a flexible and robust architecture.

\section{General Learning to Rank Frameworks}
Base on the  well-established user examination hypothesis \cite{10.1145/1242572.1242643}, we have the following equation
\begin{equation} \label{eq:1}
    P(c_d=1) = P(o_d=1)\cdot P(r_d=1)
\end{equation}
where the document will be clicked (i.e., $c_d=1$) if and only if it has been examined (i.e., $o_d=1$) and it is relevance (i.e., $r_d=1$). The goal of learning to rank is to create a ranking function $f_\theta$ that has document $\boldsymbol{d}$ as input and output a ranking score $f_\theta(d)$ that the ranked list produced by $f_\theta$ is the same as $\boldsymbol{r}$, the ranked list resulted from ranking documents by their intrinsic relevance to the query. Formally, the goal of learning to rank is to find the best $\theta$ that minimizes $f_\theta$'s loss function $\mathcal{L}$.
\begin{equation}
    \theta^{*}=\underset{\theta}{\arg \min } \mathcal{L}(\theta)=\underset{\theta}{\arg \min } \int_{q} l\left(f_{\theta}, \boldsymbol{r}_{q}\right) d P(q)
\end{equation}
where $l\left(f_{\theta}, \boldsymbol{r}_{q}\right)$ is the local ranking loss derived from the ranked list of documents produced by the learning algorithm and their relevance in each query session. $l\left(f_{\theta}, \boldsymbol{r}_{q}\right)$ is computed as 
\begin{equation} \label{eq:2}
    l(f_{\theta},\textbf{r}) = \Delta(d,r_d|f_{\theta})
\end{equation}
where $\Delta(d,r_d|f_{\theta})$ computes the individual loss on each relevant document for a ranking model $f_{\theta}$.\\
However, as stated in section 1, it is often difficult and expensive to collect relevance labels $\boldsymbol{r}$. Thus, we usually conduct learning to rank with noisy implicit feedback that correlate to relevance such as user clicks and optimize the ranking model with parameter $\theta^\dagger$ as follow.
\begin{equation}
    \theta^{\dagger}=\underset{\theta}{\arg \min } \mathcal{L}^{\prime}(\theta)=\underset{\theta}{\arg \min } \int_{q} \int_{\pi_{q}} \mathbb{E}_{\boldsymbol{c}_{\boldsymbol{\pi} q}}\left[l^{\prime}\left(f_{\theta}, \boldsymbol{c}_{\pi_{q}}\right)\right] d P\left(q, \pi_{q}\right)
\end{equation}
where $\pi_{q}$ is the ranked list achieved in the session of query $q$, $\boldsymbol{c}$ is the observed implicit feedback signal (e.g., clicks), and $ P\left(q, \pi_{q}\right)$ is the probability of displaying $\pi_{q}$ when query $q$ is submitted. As such, the task of unbiased learning to rank is to find the optimal parameters so that $\mathcal{L}(\theta) = \mathcal{L}(\theta^\dagger)$.

\section{Background}
In this section, we briefly introduce the general learning to rank frameworks as well as the algorithms currently presented in our toolbox. For simplicity, we use the standard ranking scheme where given an input query, the system retrieves and displays documents sequentially according to the order of relevance. 

Broadly speaking, there are two approaches to this task. The first approach attain unbiasness through directly debiases the implicit feed back in ranking loss's computation. The algorithms implementing this approach are called counterfactual learning algorithms. The second approach achieve unbiasness through manipulating $P\left(q, \pi_{q}\right)$ distribution. Algorithms belong to this approach is called bandit learning algorithms.

\subsection{Counterfactual Learning}
The motivation behind counterfactual learning is to remove the inherited bias of data while computing ranking loss $l^{\prime}\left(f_{\theta}, c_{\pi_{q}}\right)$ in order for the model trained with biased data a (i.e., clicks) to converge to the model trained unbiased data (i.e., the intrinsic relevance of a document). Thus, the loss function of counterfactual learning algorithms is computes as follow.
\begin{equation} \label{eq:offline_loss}
\begin{split}
    \mathcal{L}^{\prime}(\theta)& =\int_{q} \int_{\pi_{q}} \mathbb{E}_{c_{n q}}\left[l^{\prime}\left(f_{\theta}, c_{\pi_{q}}\right)\right] d P\left(q, \pi_{q}\right)\\
    &=\int_{q} \mathbb{E}_{c}\left[l^{\prime}\left(f_{\theta}, c\right)\right] d P(q)
\end{split}
\end{equation}
where $\boldsymbol{c}$ is the observed click in each session. As seen from equation \ref{eq:offline_loss}, since counterfactual learning doesn't involve the results displacement distribution in each query session (i.e., $P\left(q, \pi_{q}\right)$), it can be trained with historical data where the logging systems' distribution may not be available. Thus, most counterfactual unbiased learning to rank studies utilize search log and offline learning \cite{10.1145/3331184.3331202, 10.1145/3209978.3209986, 10.1145/3018661.3018699, 10.1145/2911451.2911537, 10.1145/3159652.3159732}.

\indent Based on equation \ref{eq:offline_loss}, we briefly describe four counterfactual algorithms that are currently supported by ULTRA. They are  the Inverse Propensity Weighting \cite{10.1145/3018661.3018699, 10.1145/2911451.2911537},the Dual Learning Algorithm \cite{10.1145/3209978.3209986}, the Regression-based EM \cite{10.1145/3159652.3159732}, and the Pairwise Debiasing model \cite{10.1145/3308558.3313447}. 
\subsubsection{\textit{Inverse Propensity Weighting}}
Inverse Propensity Weighting (IPW) updates the $l(f_0,r)$ with user feedback data $c$ as 
\begin{equation}
l_{IPW}(f_{\theta},c)=\sum_{d,c_d=1} \frac{\Delta(d,c_d|f_{\theta})}{P(O_d=1)}
\end{equation}
where $\Delta(d,c_d|f_{\theta})$ is a function that computes the individual loss on each relevant document for a
ranking model $f_{\theta}$ and $P(O_d=1)$ is the probability of document $d$ being examined in the search session. IPW estimates the examination properties by using online result randomization \cite{10.1145/2911451.2911537,10.1145/3018661.3018699} to shuffle the documents in each query randomly in order for relevance documents to have equal probabilities to be placed at each rank lists' position.
\subsubsection{\textit{Duel Leaning Algorithm}}
In order to avoid online result randomization, which affects user's experience, Duel Learning Algorithm (DLA) \cite{10.1145/3209978.3209986} simultaneously trains a ranking model $f_{\theta}$ and a examination propensity estimation model $\phi$ with an inverse relevance weight loss function (IRW) as
\begin{equation}
    l_{IRW}(\phi,c)=\sum_{d,c_d=1}\frac{\Delta(d,c_d|\phi)}{P(r_d=1)}
\end{equation}
\subsubsection{Regression EM} The Regression EM model (REM) \cite{10.1145/3159652.3159732} is another unbiased learning algorithm that does not depend on result randomization. It achieves this by using utilizing a graphic model in correspondence with an EM algorithm to unify ranking model training and examination propensity estimating. Using the user examination hypothesis described in Eq. (\ref{eq:1}), the computation for likelihood of observed clicks for each query $q$ in REM is
\begin{equation}\scriptstyle
    \log P(c) = \sum_d c_d\log(P(o_d=1)\cdot P(r_d=1)) + (1-c_d)\log(1-P(o_d=1)\cdot P(r_d=1)
\end{equation}
where $c_d$ is extracted from online user interactions or search log, $o_d$ and $r_d$ are latent variables and $P(r_d=1)$ is calculated based on ranking function $f_{\theta}$ as 
\begin{equation}
    P(r_d=1) = \frac{1}{1 + \exp(-f_{\theta}(d))}
\end{equation}
\subsubsection{Pairwise Debiasing.} The Pairwise Debiasing (PD) model, proposed by Hu et al. \cite{10.1145/3308558.3313447}, conducts learning to rank using inverse propensity weighting. Similar to DLA, PD also trains examination propensity estimation models together with the ranking
models. However, there are two main differences between the two algorithms. First, PD is specifically designed to use with pairwise learning to rank where $l(f_{\theta}, \textbf{r})$ is computed as the sum of pairwise losses $\Delta(f_{\theta},d^+,d^-)$ where $r_{d^+} > r_{d^-}$. Second, PD also take into account of unclicked documents during training process by assuming \cite{10.1145/3308558.3313447}
\begin{equation}
    P(c_d=0) = t\cdot P(r_d=1)
\end{equation}
PD's inverse propensity weighted version of $l(f_{\theta},r)$ is computed as 
\begin{equation}
    l_{PD}(f_{\theta},c)=\sum_{d^+,d^-,c_{d^+}=1, c_{d^-}=0} \frac{\Delta(f_{\theta},d^+,d^-)}{P(O_d=1)\cdot t}
\end{equation}
where $P(O_d=1)$ and $t$ is estimated in a similar fashion as REM.
\subsection{Bandit Learning Algorithms}
The idea behind bandit learning is update the model by analyzing observed real-time user feedback in controlled environment. With unbiased learning to rank, this means manipulating the ranked lists for each query session and estimating unbiased gradients from click data. The loss function $\mathcal{L}^\prime$ is computed as follow.
\begin{equation}
    \mathcal{L}^{\prime}(\theta)=\int_{q} \int_{\pi_{q}} \mathbb{E}_{\boldsymbol{c}_{\boldsymbol{\pi}_{q}}}\left[l\left(f_{\theta}, \boldsymbol{c}_{\pi_{q}}\right)\right] d P\left(\pi_{q} \mid q\right) d P(q)
\end{equation}
Bandit learning algorithms achieve unbiasness by finding $P\left(\pi_{q} \mid q\right)$ so that for any $(\theta, \theta^\prime)$:

$\begin{aligned} \int_{\pi_{q}} \mathbb{E}_{c_{\pi q}}\left[l\left(f_{\theta}, \boldsymbol{c}_{\pi_{q}}\right)\right] d P\left(\pi_{q} \mid q\right) & \geq \int_{\pi_{q}} \mathbb{E}_{c_{\pi q}}\left[l\left(f_{\theta^{\prime}}, \boldsymbol{c}_{\pi_{q}}\right)\right] d P\left(\pi_{q} \mid q\right) \\ \Rightarrow l\left(f_{\theta}, \boldsymbol{r}\right) & \geq l\left(f_{\theta^{\prime}}, \boldsymbol{r}\right) \end{aligned}$
\\
Since bandit learning algorithms requires the manipulation of\\ $P\left(\pi_{q} \mid q\right)$, they are usually implemented in online environment. As a result, they are commonly referred as \textit{online learning to rank algorithms}.

In the following section, we discuss the four bandit learning algorithms that are available in ULTRA, which are the Dueling Bandit Gradient
Descent \cite{10.1145/1553374.1553527}, Multileave Gradient Descent \cite{10.1145/2835776.2835804}, Null Space Gradient Descent \cite{10.1145/3209978.3210045}, and the Pairwise Differentiable Gradient Descent algorithm \cite{Oosterhuis_2018}.
\subsubsection{Dueling Bandit Gradient Descent.} The Dueling Bandit Gradient Descent model, proposed by Yue and Joachims \cite{10.1145/1553374.1553527}, performs $f_{\theta}$ optimization in three steps:
\begin{itemize}
    \item \textit{Step 1}: Generate parameter $\theta '$ by sampling sampling parameter perturbation and adding it to the original parameter $\theta$. This effectively makes ranked list $\pi_\theta$ and $\pi_{\theta'}$, produced by $f_\theta$ and $f_{\theta'}$ correspondingly, different.
    \item \textit{Step 2}: Compute the two losses $l(f_\theta, c_{\pi_\theta}$ and $l(f_{\theta'}, c_{\pi_\theta'}$ by collecting click from showing $\pi_\theta$ and $\pi_{\theta'}$ (directly or interleavedly) to real user.
    \item \textit{Step 3}: If we observe $\EX_{c_{\pi_\theta'}}[l(f_{\theta'}, c_{\pi_\theta'}] <$ $\EX_{c_{\pi_\theta}}[l(f_{\theta}, c_{\pi_\theta}]$
    in online experiment then update $\theta$ with $\theta'$
\end{itemize}
By repeating these steps, we can achieve unbiased ranking model.
\subsubsection{Multileave Gradient Descent.} Multileave Gradient Descent (MGD) \cite{10.1145/2835776.2835804} is an extension of DBGD. Similar to DBGD, MGD also shares the three steps optimization. However, instead of sampling the parameter perturbation $\theta'$ once every train step, MGD samples multiple $\theta'$ and compares the set of ranked list produced for a better candidates selection.
\subsubsection{Null Space Gradient Descent.} Just like MGD, Null Space Gradient Descent (NSGD) \cite{10.1145/3209978.3210045} is also an extension of DBGD that use multiple parameter perturbations $\theta'$. The difference is that NSGD stores explored perturbed parameters that resulted in poorly perform gradient in previous training instances and sample new parameters from null space for more efficient direction exploration.
\subsubsection{Pairwise Differentiable Gradient Descent.} Pairwise Differentiable Gradient Descent (PDGD) \cite{Oosterhuis_2018} break away from using interleaved and multileaved as well as extensive model sampling by applying the Pluckett-Luce model to $f_\theta$ to create a distribution over $\boldsymbol{q}$. Then PDGD stochastically samples ranked list $\pi_q$ as
\begin{equation}
P\left(\pi_{q} \mid q\right)=\prod_{i=1}^{\left|\pi_{q}\right|} \frac{\exp \left(f_{\theta}\left(d_{i}\right)\right)}{\sum_{j=i}^{\left|\pi_{q}\right|} \exp \left(f_{\theta}\left(d_{j}\right)\right)}
\end{equation}
where $d_i$ is the ith document in $\pi_{q}$. PDGD computes the $l(f_\theta, \boldsymbol{c})$ by summing pairwise losses $\Delta(f_\theta,d_i,d_j)$ over document pairs as
\begin{equation}
    \Delta^{\prime}\left(f_{\theta}, \boldsymbol{c}_{\pi_{q}}\right)=\sum_{d_{i}, d_{j}, j<i+2, c_{d_{i}}=1, c_{d_{j}}=0} \rho\left(d_{i}, d_{j}, \pi_{q}\right) \cdot \Delta\left(f_{\theta}, d_{i}, d_{j}\right)
\end{equation}
where $d_i$ is the clicked document under $d_j$ in $\pi_{q}$ and $\rho\left(d_{i}, d_{j}, \pi_{q}\right)$ is computed as 
\begin{equation}
    \rho\left(d_{i}, d_{j}, \pi_{q}\right)=\frac{P\left(\pi_{q}\left(d_{i}, d_{j}\right) \mid q\right)}{P\left(\pi_{q} \mid q\right)+P\left(\pi_{q}\left(d_{j}, d_{i}\right) \mid q\right)}
\end{equation}
where $\pi_{q}(d_{j}, d_{i})$ is the ranked list with the documents in position $i$ and $j$ reversed.

\section{Architecture}
The purpose of our ULTRA toolbox is to provide a code base for experiment and research on Unbiased Learning to Rank algorithms using different datasets. Our toolbox is built using Pytorch framework which is easy to use and expanded upon. Using the unified data processing pipeline, ULTRA support multiple unbiased counterfactual and bandit learning algorithms and ranking models. We also developed different classes to simulate noisy labels that help with training and testing different algorithms and ranking models.
\subsection{ULTRA's packages}
There are four packages in ULTRA namely \textbf{Input Layer}, \textbf{Learning Algorithms}, \textbf{Ranking Models}, and \textbf{Utils}. UTRA is divided into four package so that users easily can pick which click simulators, algorithms, models, and evaluation metrics. It is like play Lego where users can just swap out different building blocks and they can achieve different results. Moreover, by separating the toolbox into different packages, we also make sure that the implementation of one process doesn't affect another process. For example, if a researcher is to add in a new learning algorithm, they can still utilize the available ranking model, click simulation and evaluation metrics as long as their new algorithm follow the same structure as the previous one. Most packages has a base module that can be used as a framework so that adding in new algorithms is easy and remain consistent with the current structure. 
\subsubsection{Input Layer.} Input Layer main functions are processing the data and simulate users' clicks to feed into the Learning Algorithm Layers. All the classes in this layer maintain a reference to the downstream ranking model to help with the creation of the input feed. This layer supports four algorithms namely \textbf{Direct Label Feed}, \textbf{Click Simulation Feed}, \textbf{Deterministic Online Simulation Feed} and \textbf{Stochastic Online Simulation Feed}. The Input Layer has a \texttt{base\_input\_feed} module and all the supported algorithms in this layer is an instance of it. As a result, the base module consists of three main abstract methods that are shared among all algorithms. 
\begin{itemize}
    \item \texttt{preprocess\_data(data\_set,hparam\_str,exp\_settings)} is called before training or testing to preprocess the data based on the input feed. 
    \item \texttt{get\_batch(data\_set, check\_validation)} is called every training iteration to get a random batch of data and prepare the input feed to train the model. The input feed consist of document feature array, label arrays, and document id arrays. The number of label arrays and document id arrays is equal to the number of ranking documents to be considered in each query. If an online simulation module is used, then the input feed also consists of the winning ranking models' results.
    \item \texttt{get\_next\_batch(data\_set, check\_validation)} is called during validation step to get data from specifics index. It is to make sure that we iterate through all the data in the validation set.
\end{itemize}
This abstract methods allow users' to add their own click simulations while remain consistent with the current API structure. Each sub module of this base module has their own methods to assist with creating the input feed for the learning to rank model. For offline learning, the class \texttt{click\_simulation\_feed} has an important helper function.
\begin{itemize}
    \item \texttt{prepare\_sim\_clicks\_with\_index(data\_set,index,docid\_inputs,\\letor\_features,labels,check\_validation)} is called to simulate click based on a pre-defined click model (described with section 4.1.4) to create click labels for the ranking model. The function returns the document ids in each ranking position, the click labels and the feature of each document. 
\end{itemize}
The direct label feed, deterministic online simulation and stochastic online simulation have a different helper function.
\begin{itemize}
    \item \texttt{prepare\_true\_labels\_with\_index(data\_set,index,docid\_inputs,\\letor\_features,labels,check\_validation)} is used inside the \texttt{get\_batch} and \texttt{get\_next\_batch} functions to prepare the document relevance label, ids and feature for the dataset. It basically converts all the negative relevance labels in the raw data to 0. 
\end{itemize}

The deterministic and stochastic online click simulations have another helper function.
\begin{itemize}
    \item \texttt{simulate\_clicks\_online(input\_feed,check\_validation)} is \\called inside the \texttt{get\_batch} and \texttt{get\_next\_batch} functions to simulate online clicks. For deterministic online simulation, the function first creates a ranked list by sorting the documents according to the output of the current ranking models and then generate synthetic clicks. For stochastic online simulation, the function simulate synthetic clicks by sampling documents from a distribution generates from using Plackett-Luce model over the current ranking model output. 
\end{itemize}
\subsubsection{Learning Algorithms Package.} Learning Algorithms Layer initialize and train the ranking model using the data from the Input Layer and evaluate the result based on chosen metrics. This package is designed with a \texttt{base\_algorithm} module and eight sub modules that are instances of the base one. This design allows researchers to easily add their own learning to rank algorithm. The \texttt{base\_algorithm} currently provides four loss functions that can be used interchangeably to fit users' requirements
\begin{itemize}
    \item \texttt{pairwise\_cross\_entropy\_loss(pos\_scores,neg\_scores,\\propensity\_weights)} is used to computes the pairwise softmax loss.
    \item \texttt{sigmoid\_loss\_on\_list(output,labels,propensity\_weights)} is \\used to compute the pointwise sigmoid loss.
    \item \texttt{pairwise\_loss\_on\_list(output,labels,propensity\_weights)} is used to compute the pairwise entropy loss.
    \item \texttt{softmax\_loss(output,labels,propensity\_weights)}is used to compute the listwise softmax loss.
\end{itemize}
Beside these loss functions, this module also provides abstract methods and helper functions for training and updating the ranking models.
\begin{itemize}
    \item \texttt{train(input\_feed)} is an abstract method that is called to perform an one step of training the model. 
    \item \texttt{validation(input\_feed)} is an abstract method that is called to validate the ranking model or when the input layer needs to create a ranked list to simulate online clicks.
    \item \texttt{remove\_padding\_for\_metric\_eval(input\_id\_list,model\_output)} is called to remove the padding of the ranking model's output for evaluation in each training and validating steps. 
    \item \texttt{create\_model(feature\_size)} is called at the beginning of training to initialize the ranking model with document's feature size.
    \item \texttt{ranking\_model(model,list\_size)} is called during each training and validating step to perform a forward pass of the model and get the model's output. 
    \item \texttt{create\_input\_feed(input\_feed,list\_size)} is called before each training and validating step to generate the input for the ranking model and the labels for evaluating model's output. 
    \item \texttt{create\_summary(scalar\_name,summarize\_name,value,\\is\_training)} is during training and validating to log important values such as iteration loss, evaluation scores, propensity weights, etc.
    \item \texttt{opt\_step(opt,params)} is called at the end of each training step to perform gradient update.
\end{itemize}
These functions make adding or changing algorithms more smoothly and with little code repetition. They could be override according to users' needs. All the functions and abstract methods in the base module make sure that all the algorithms is created using the same structure, and input while being evaluated with same the loss functions and metrics. It is important for these algorithms to have a unify way of producing unbiased model so that they can be compared and benchmark. 

Currently, ULTRA supports \textbf{DLA}, \textbf{IPW}, \textbf{REM}, \textbf{PD}, \textbf{DBGD}, \textbf{MGD}, \textbf{NSGD} and \textbf{PDGD}. Each of this algorithm is a sub module of the \texttt{base\_algorithm} with a different implementations of the \texttt{train} and \texttt{validation} to calculate the model's result and loss. The evaluation metrics for these models are invoked from the \textbf{Utils} package. The separation of these two packages means that the evaluation process are consistent regardless of the implementation of \texttt{train} and \texttt{validation} functions.
\subsubsection{Ranking Models.} This layer contains the code for different ranking models that is initialized within the Learning Algorithms Layer. Similar to the previous two layers, this layer contains a \texttt{base\_ranking\_model} module that currently has two sub modules \texttt{DNN} and \texttt{Linear}. There is one method.
\begin{itemize}
    \item \texttt{build(input\_list, noisy\_params,noise\_rate,is\_training,\\**kwargs)} is called from the Learning Layer to do a forward pass with the input list and return an output.
\end{itemize}
Each sub module has their own version of build as well as different model structures. The \texttt{Linear} module uses a linear function to compute ranking score. The \texttt{DNN} module computes ranking scores using a multi-layer perceptron network with a non-linear activation function. Both of these module implement a normalization technique on the input before processing them. 
\subsubsection{Utils Layer.} The Util Layer has different sub modules to help with other layers in different tasks. As a result, it does not have a base module. For simplicity sake, we will not go into detail of each function in each sub module but rather just summarize what each them do. 
\begin{itemize}
    \item \texttt{click\_models}: This module provides implementation of the Positional Bias Model, User Browsing Model, and Cascade Model. These models are initialized and called in the Input Layer to produce synthetic clicks.
    \item \texttt{data\_utils}: This module processes the raw data into correct format for input layer to use. It is used in the main class where everything is initialized and run.
    \item \texttt{hparams}: This module is also used in the main class. It main purpose is to setup all the configuration parameters for the input layer, learning algorithm layer, and ranking models layer.
    \item \texttt{metric}: This module is invoked by the algorithm layer to evaluate the ranking results using metrics set by users. It currently supports eight ranking metrics namely Mean Reciprocal Rank, Expected Reciprocal Rank, Average Relevance Position, Discounted Cumulative Gain, Normalized Discounted Cumulative Gain, Precision, Mean Average Precision, and Order Pair Accuracy.
    \item \texttt{metric\_utils}: This module provides methods to help with metric evaluation in the \texttt{metric} module such as sorting the output or the ranking scores.
    \item \texttt{propensity\_estimator}: This module provides three propensity estimators: the Basic Propensity Estimator, the Oracle Propensity Estimator, and the Randomized Propensity Estimator. These estimators are initialized and invoked in the IPW algorithm belonged to the Learning Algorithm layer.
    \item \texttt{sys\_tool}: This module assists with initializing class instances using the configuration files created by users.
    \item \texttt{team\_draft\_interleave}: This module performs interleaving on online learning algorithms' results. It is invoked from the Deterministic and Stochastic Online simulation of the Input Layer. 
\end{itemize}
\subsection{Pipeline Life Cycle.}
\begin{figure}
    \centering
    \includegraphics[scale=0.215]{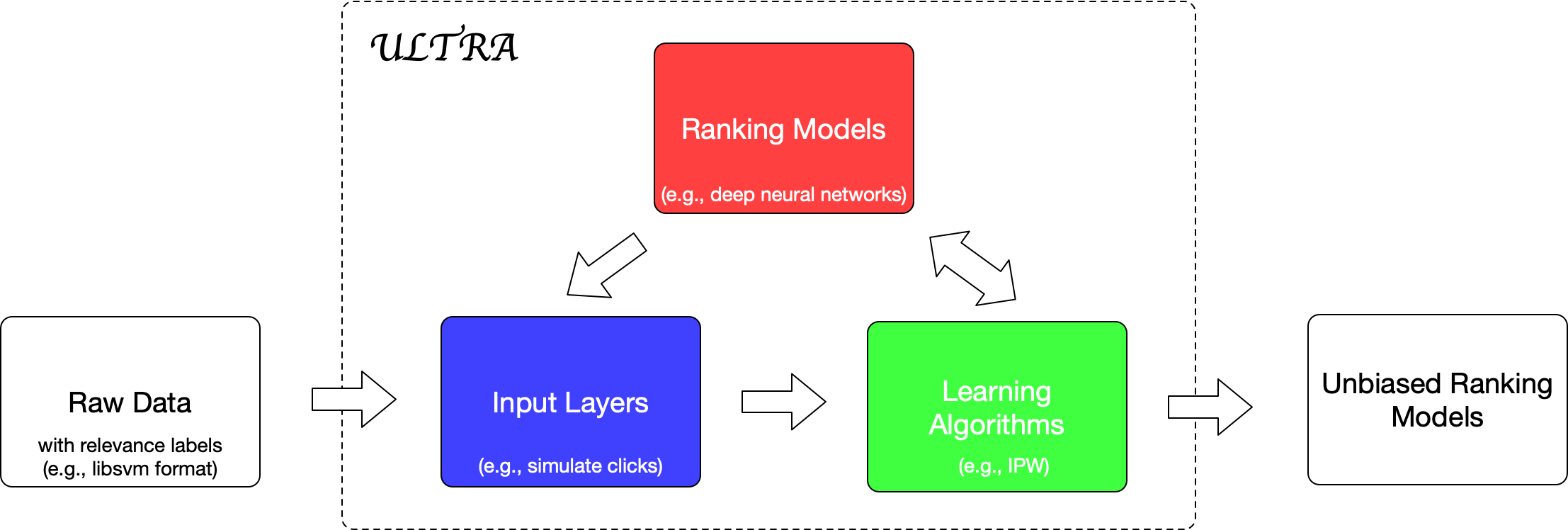}
    \caption{Raw data is piped into the input layer to be processed and simulate clicks. This is then fed into the learning algorithm to train the ranking model. Depending of which algorithms is being used, the model output ranked list is fed back into the input layer to simulate clicks online to calculate model loss. This cycle is repeated to produce unbiased ranking models.}
    \label{fig:ULTRA_pipeline}
\end{figure}

This section provides a detailed description of ULTRA pipeline. The general pipeline of ULTRA is described in figure \ref{fig:ULTRA_pipeline}. 
\begin{enumerate}
    \item Users need to first create a JSON file specifying the click simulation method, learning algorithm and evaluation metric. When the toolbox is run, it will read the Json file and initialize the learning algorithm model and ranking model in the JSON file.
    \item Raw data is read into the toolbox and processed into ULTRA format.
    \item During each of the training step, the Input Layer will get a random batch of data, and generate synthetic clicks. If the learning algorithms is a bandit learning algorithm, the Input Layer instead get a ranked list using current ranking model and generates the input feed with that list.
    \item The input feed created by the Input Layer will be fed to the learning to rank model initialized in \textit{step 1}. The output of the model will be used to calculate loss and gradient. 
    \item Repeat step (2) to (4) for how many iteration config by the users.
\end{enumerate}
\subsection{Configurations and features}
\begin{table}[]
    \centering
    \begin{tabular}{{|p{.15\textwidth}|p{.28\textwidth}|}}
    \hline 
       \makecell[l]{
       - \texttt{train\_input\_feed}\\
       - \texttt{valid\_input\_feed}\\
       - \texttt{test\_input\_feed}}  & \makecell[lc]{Click simulation algorithms for training\\validating, and testing} \\
     \hline
        \makecell[l]{
       - \texttt{train\_input\_hparams}\\
       - \texttt{valid\_input\_hparams}\\
       - \texttt{test\_input\_hparams}} & \makecell[l]{The parameters for training, validating\\and testing algorithm:\\
        - \texttt{click\_model\_json}: the setting file for\\ the predefined click models.\\
        - \texttt{oracle\_mode}: set true to feed relevance\\ labels instead of simulated clicks.\\
        - \texttt{dynamic\_bias\_eta\_change}: set eta\\ change step for dynamic bias severity \\in training, 0.0 means no change.\\
        - \texttt{dynamic\_bias\_step\_interval}: set how\\ many steps to change eta for dynamic\\ bias severity in training, 0.0 means\\ no change.}\\
        \hline
        \texttt{ranking\_model} & \makecell[lc]{Ranking model for training}\\
        \hline
        \texttt{ranking\_model\_hparams} &\makecell[l]{
        - \texttt{hidden\_layer\_sizes}: number of\\neurons in each layer of the ranking\\model\\
        - \texttt{activation\_func}: type of activation\\function, which could be elu, relu,\\sigmoid, or tanh\\
        - \texttt{norm}: type of normalization, which\\could be BatchNorm or LayerNorm}\\
        \hline
        \texttt{learning\_algorithm} & \makecell[lc]{Learning algorithm to train the model}\\
        \hline
        \texttt{learning\_algorithms\-\_hparams} &\makecell[l]{The hyper parameters for learning\\algorithms. Below are the parameters\\that are\\required among all learning algorithms\\
        - \texttt{learning\_rate}: learning rate\\
        - \texttt{max\_gradient\_norm}: clip gradients to\\this norm\\
        - \texttt{grad\_strategy}: optimizer to use for\\calculating gradient\\
        Beside these three, each learning\\algorithm has their own hyper\\parameters and all of these parameters\\have default value.}\\
        \hline
        \texttt{metrics} & A list of valuation metrics to use.\\
        \hline
        \texttt{metrics\_topn} & Number of ranking document for metric evaluation consideration\\
        \hline
        \texttt{objective\_metric} & The metric used to store the model that perform best.\\
        \hline
    \end{tabular}
    \caption{Configuration Parameters}
    \label{tab:conf_param}
\end{table}
ULTRA can be configured to produce unbiased ranking models with different learning algorithms and model structures. These
configuration parameters are described in Table \ref{tab:conf_param}.\\
Moreover, users can set several other command line arguments to train the ranking model as follow.
\begin{itemize}
    \item \texttt{data\_dir}: the directory of the dataset.
    \item \texttt{train\_data\_prefix}: The name prefix of the training data in \texttt{data\_dir}.
    \item \texttt{valid\_data\_prefix}: The name prefix of the validation data in \texttt{data\_dir}.
    \item \texttt{test\_data\_prefix}: The name prefix of the test data in  \texttt{data\_dir}.
    \item \texttt{model\_dir}: The directory for model and intermediate outputs.
    \item \texttt{output\_dir}: The directory to output results.
    \item \texttt{setting\_file}: A json file that contains all the settings of the algorithm.
    \item \texttt{batch\_size}: Batch size to use during training.
    \item \texttt{max\_list\_cutoff}: The maximum number of top documents to consider in each rank list (0: no limit).
    \item \texttt{selection\_bias\_cutoff}: The maximum number of top documents to be shown to user (which creates selection bias) in each rank list (0: no limit).
    \item \texttt{max\_train\_iteration}: Limit on the iterations of training (0: no limit).
    \item \texttt{start\_saving\_iteration}: The minimum number of iterations before starting to test and save models (0: no limit).
    \item \texttt{steps\_per\_checkpoint}: How many training steps to do per checkpoint.
    \item \texttt{test\_while\_train}: Set to True to test models during the training process.
    \item \texttt{test\_only}: Set to True for testing models only.
\end{itemize}
Users can also configure different click simulation models from command line using the arguments as follow.
\begin{itemize}
    \item Click model name. Users can choose among position bias model (\texttt{pbm}), cascade model (\texttt{cascade}), and user browsing model (\texttt{ubm}).
    \item Negative click probability is the probability that a document is examine but not clicked based on the relevance. 
    \item Positive click probability is the probability that a document is examine and clicked based on the relevance. 
    \item Max relevance grade what is the maximum score of relevancy that a document could achieve.
    \item the learning rate (eta)
\end{itemize}
A detailed description on how to create a click model 

\section{Use Case}
To evaluate our toolbox, we created two ranking models using IPW and DBGD algorithms with ULTRA.
\subsection{IPW learning algorithm}
\begin{figure}[!hbt]
    \centering
    \includegraphics[scale=0.64]{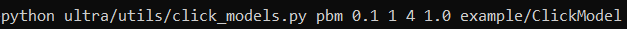}
    \caption{Creating Click Model with ULTRA using Command line}
    \label{fig:clickmodel}
\end{figure}

\begin{figure}[!hbt]
    \centering
    \includegraphics[scale=0.52]{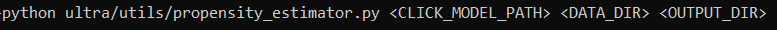}
    \caption{Propensity Estimator with ULTRA using Command line}
    \label{fig:PEstimator}
\end{figure}

First, we created the Positional Biased Click Model with ULTRA. This is the model that simulate users' position bias.  We set the negative click probability, positive click probability, max relevance grade and eta as 0.1, 1, 4, 1.0 respectively. We put the model \texttt{pbm\_0.1\_1.0\_4\\\_1.0.json} file and saved it in \texttt{./examp\\le/ClickModel/}. The command line for creating the click model is shown in figure \ref{fig:clickmodel}
Next, we created our Randomized Propensity Estimator for the IPW algorithms. This is done using the command line through ULTRA as described in figure \ref{fig:PEstimator}.
Then in the setting file, we choose Click Simulation Feed algorithm to create the input feed for training since IPW is a counterfactual learning algorithm. For validation and testing, we instead choose Direct Label Feed algorithm to create the input feed for the model since our goal is for the model to correctly rank the document according to relevance. The  values for the train, valid and test hyper parameters are as follow.
\begin{itemize}
    \item \texttt{oracle\_mode = False}
    \item \texttt{dynamic\_bias\_eta\_change=0.0}
    \item \texttt{dynamic\_bias\_step\_interval=1000}
    \item \texttt{click\_model\_json} to \texttt{./example/ClickModel\\/pbm\_0.1\_1.0\_4\_1.0.json}, our click model json file that we saved earlier .
\end{itemize}
For the ranking model, we choose the standard multi-layer perceptron network with \texttt{hidden\_layer\_sizes= [512, 256, 128]}. We use pytorch's LayerNorm \cite{ba2016layer} as our normalization layer, and activation function to be the ELU function which is defined as
\begin{equation}
    E L U(x)=\left\{\begin{array}{ll}
x, & \text { if } x \geq 0 \\
e^{x}-1, & \text { if } x<0
\end{array}\right.
\end{equation}
For the learning algorithm, we choose IPW with the following hyper parameters.
\begin{itemize}
    \item \texttt{propensity\_estimator\_type}: the Randomized Propensity Estimator
    \item \texttt{propensity\_estimator\_json}: the Randomized Propensity Estimator json file that we created.
    \item \texttt{learning\_rate}: 0.05.
    \item \texttt{max\_gradient\_norm}: 5.0.
    \item \texttt{loss\_function}: softmax loss
    \item \texttt{l2\_loss}: 0.0
    \item \texttt{grad\_strategy}: Ada Grad
\end{itemize}
Users can easily change them to fit their requirement. For evaluation metrics, we choose NDCG and ERR. Our test data is the Yahoo! dataset. We set our training iteration to be 10,000, batch size to be 256, selection bias cutoff to be 10. We achieved our ranking model after running the toolbox.

\subsection{DBGD learning algorithm}
Since we already created our click model to run IPW, we can reuse that model to create our input feed for DBGD thanks to ULTRA high reusability. For the setting file, instead of choosing Click Simulation Feed as our input layer for training, we choose Stochastic Online Simulation Click since it works better with online learning algorithm such as DBGD. We keep the ranking model the same for DBGD. For learning algorithm, we set the following hyper parameters.
\begin{itemize}
    \item \texttt{learning\_rate}: 0.05.
    \item \texttt{max\_gradient\_norm}: 5.0.
    \item \texttt{need\_interleave}: True
    \item \texttt{grad\_strategy}: Ada Grad
\end{itemize}
Our evaluation metrics, test data and command line arguments are the same as IPW's.
\subsection{Observation}
As we can see from these two use cases, experimenting on ULTRA with different algorithms is just as simple as changing the name of the algorithm. ULTRA's API structure allow users to test algorithms with different settings with ease. Moreover, its high reusability means that users can save a lot of time by not having to recreate click model for every experiment.
\begin{table*}[t]
    \caption{\textmd{Results of unbiased learning-to-rank (ULTR) algorithms using ULTRA toobox with different learning paradigms on Yahoo! LETOR data using multi-layer perceptron as ranking models.}}
    \subcaption*{\textmd{(a) Performance of unbiased learning-to-rank algorithms with offline learning on Yahoo! LETOR data}}
    \centering
    \begin{tabular}{p{2.25cm}|c||c|c|c|c|c|c|c|c}
    \hline
    \multicolumn{10}{c}{Offline Learning}\\
    \hline
    \hline
    \multicolumn{2}{c||}{} & nDCG@1 & ERR@1 & nDCG@3 & ERR@3 & nDCG@5 & ERR@5 & nDCG@10 & ERR@10  \\
    \hline
    \hline
    \multirow{4}{10 em}{Counterfactual\\ Learning Family}& IPW & 0.695\
 & 0.344 & 0.706 & 0.425 & 0.709 & 0.447 & 0.745 & 0.472  \\
    \cline{2-10}
    & REM & 0.657 & 0.341 & 0.665 & 0.416 & 0.677 & 0.432 & 0.722 & 0.454  \\
    \cline{2-10}
    & DLA & 0.701 & 0.351 & 0.708 & 0.422 & 0.711 & 0.439 & 0.756 & 0.463  \\
    \cline{2-10}
    & PairD & 0.681 & 0.319 & 0.706 & 0.407 & 0.714 & 0.435 & 0.732 & 0.447  \\
    \hline
    \hline
    \multirow{4}{10 em}{Bandit Learning\\ Family}& DBGD & - & - & - & - & - & - & - & -  \\
    \cline{2-10}
    & MGD & - & - & - & - & - & - & - & -  \\
    \cline{2-10}
    & NSGD & - & - & - & - & - & - & - & -  \\
    \cline{2-10}
    & PDGD & 0.319 & 0.117 & 0.372 & 0.189 & 0.410 & 0.230 & 0.521 & 0.249  \\
    \hline
    \hline
    \end{tabular}
    \subcaption*{\textmd{(b) Performance of unbiased learning-to-rank algorithms with stochastic online learning on Yahoo! LETOR data}}
    \centering
    \begin{tabular}{p{2.25cm}|c||c|c|c|c|c|c|c|c}
    \hline
    \multicolumn{10}{c}{Stochastic Online Learning}\\
    \hline
    \hline
    \multicolumn{2}{c||}{} & nDCG@1 & ERR@1 & nDCG@3 & ERR@3 & nDCG@5 & ERR@5 & nDCG@10 & ERR@10  \\
    \hline
    \hline
    \multirow{4}{10 em}{Counterfactual\\ Learning Family}& IPW & 0.684 & 0.348 & 0.709 & 0.427 & 0.725 & 0.446 & 0.758 & 0.475  \\
    \cline{2-10}
    & REM & 0.687 & 0.345 & 0.689 & 0.416 & 0.694 & 0.435 & 0.739 & 0.455  \\
    \cline{2-10}
    & DLA & 0.673 & 0.352 & 0.681 & 0.420 & 0.692 & 0.436 & 0.745 & 0.465  \\
    \cline{2-10}
    & PairD & 0.674 & 0.322 & 0.715 & 0.410 & 0.725 & 0.437 & 0.740 & 0.446  \\
    \hline
    \hline
    \multirow{4}{10 em}{Bandit Learning\\ Family}& DBGD & 0.413 & 0.166 & 0.474 & 0.259 & 0.521 & 9.291 & 0.614 & 0.308  \\
    \cline{2-10}
    & MGD & 0.426 & 0.172 & 0.478 & 0.270 & 0.541 & 0.312 & 0.614 & 0.326  \\
    \cline{2-10}
    & NSGD & 0.435 & 0.179 & 0.496 & 0.273 & 0.539 & 0.305 & 0.621 & 0.327  \\
    \cline{2-10}
    & PDGD & 0.687 & 0.352 & 0.691 & 0.430 & 0.711 & 0.452 & 0.754 & 0.461  \\
    \hline
    \hline
    \end{tabular}
    \subcaption*{\textmd{(c) Performance of unbiased learning-to-rank algorithms with deterministic online learning on Yahoo! LETOR data}}
    \centering
    \begin{tabular}{p{2.25cm}|c||c|c|c|c|c|c|c|c}
    \hline
    \multicolumn{10}{c}{Deterministic Online Learning}\\
    \hline
    \hline
    \multicolumn{2}{c||}{} & nDCG@1 & ERR@1 & nDCG@3 & ERR@3 & nDCG@5 & ERR@5 & nDCG@10 & ERR@10  \\
    \hline
    \hline
    \multirow{4}{10 em}{Counterfactual\\ Learning Family}& IPW & 0.682 & 0.347 & 0.706 & 0.427 & 0.724 & 0.445 & 0.756 & 0.473  \\
    \cline{2-10}
    & REM & 0.671 & 0.343 & 0.708 & 0.417 & 0.716 & 0.433 & 0.738 & 0.454  \\
    \cline{2-10}
    & DLA & 0.685 & 0.352 & 0.713 & 0.421 & 0.728 & 0.437 & 0.754 & 0.464  \\
    \cline{2-10}
    & PairD & 0.673 & 0.320 & 0.716 & 0.409 & 0.724 & 0.436 & 0.740 & 0.447  \\
    \hline
    \hline
    \multirow{4}{10 em}{Bandit Learning\\ Family}& DBGD & 0.359 & 0.136 & 0.421 & 0.223 & 0.471 & 0.253 & 0.571 & 0.280  \\
    \cline{2-10}
    & MGD & 0.419 & 0.175 & 0.479 & 0.265 & 0.532 & 0.296 & 0.609 & 0.322  \\
    \cline{2-10}
    & NSGD & 0.426 & 0.159 & 0.480 & 0.255 & 0.528 & 0.291 & 0.616 & 0.314  \\
    \cline{2-10}
    & PDGD & 0.661 & 0.342 & 0.673 & 0.421 & 0.685 & 0.441 & 0.742 & 0.459  \\
    \hline
    \hline
    \end{tabular}
    \label{tab:ULTRA_result}
\end{table*}
\section{Experiments and results}
In this section, we experimented with all our current learning algorithms using the same hyper parameters and ranking model as described in the use cases. We performed three different experiments, one with Click Simulation Feed just for the counterfactual learning algorithms and PDGD, one for all algorithms with deterministic online learning simulation and one with stochastic online learning simulation. We trained and tested the models with the predefined training, validation, and test data in Yahoo!. We used NDCG as our ranking metrics for evaluation. We repeated the experiment 5 times and average the metric values on top 1, 3, 5 and 10 results. The result of the experiment is reported in Table \ref{tab:ULTRA_result}. As we can see from the result, IPW, DLA and PDGD perform the best among all the algorithms that were tested. In offline learning, IPW and DLA achieve the best nDCG@10. In online learning, IPW and DLA perform the best for Deterministic Online Learning while PDGD and IPW achieve best performance in Stochastic Online Learning. When comparing offline and online learning performance for counterfactual learning family, we do not notice any significant differences in the results. The result in this paper align with the result from experiments conducted by Ai et al. \cite{10.1145/3439861}.
\section{Availability}
\begin{figure}[!hbt]
    \centering
    \includegraphics[scale=0.82]{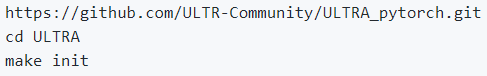}
    \caption{Step to setup ULTRA}
    \label{fig:ULTRA_setup}
\end{figure}

ULTRA is available on GitHub at https://github.com/ULTR-\\Community/ULTRA\_pytorch under the Apache License. To run the simulator, the repository should be cloned locally and installed all the necessary dependencies as shown in Figure \ref{fig:ULTRA_setup}. The json setting file contains parameters as described in Section 4.3 that allow customization in the learning to rank process. \texttt{ULTRA/pytorch/example/\\Yahoo} provides example on how to train ranking model with 1000 iteration on Yahoo! dataset. The \texttt{ULTRA/pytorch/example} folder also contains other example and setting to create different learning to rank models as well as premade click simulation model and propensity estimator. For reproductibility purpose, all the hyper parameters setting in section 6 is used a default value.

\section{Conclusion and Future Work}
In this paper, we presented ULTRA toolbox, a code base for experimenting and researching learning to rank algorithms. The primary features of
our toolbox include flexible architecture for testing learning to rank algorithms with different settings, tunable parameters for click simulation feed, learning algorithms, ranking models and evaluation metrics, and extensible API for expanding the toolbox with users' personal algorithms or models.

We detailed two use cases to generate unbiased ranking models using IPW algorithms and DBGD algorithm. We proceeded to evaluate our toolbox by running multiple experiments on all of our current learning to rank algorithms with different click simulations. ULTRA is highly customizable and users can easily configure parameters such as setting different learning rate, click models, optimization strategy, and other model training parameters.

For future work, we plan to adding more learning to rank algorithms and ranking models. We will also aim to optimize the toolbox to improve the execution speed.

\section{ACKNOWLEDGMENTS}
This work was supported by the School of Computing, University of Utah. Any opinions, findings and conclusions or recommendations expressed in this material are those of the authors and do not necessarily reflect those of the sponsor.

\bibliographystyle{ACM-Reference-Format}
\bibliography{ULTRA}

%%% -*-BibTeX-*-
%%% Do NOT edit. File created by BibTeX with style
%%% ACM-Reference-Format-Journals [18-Jan-2012].

\begin{thebibliography}{35}

%%% ====================================================================
%%% NOTE TO THE USER: you can override these defaults by providing
%%% customized versions of any of these macros before the \bibliography
%%% command.  Each of them MUST provide its own final punctuation,
%%% except for \shownote{}, \showDOI{}, and \showURL{}.  The latter two
%%% do not use final punctuation, in order to avoid confusing it with
%%% the Web address.
%%%
%%% To suppress output of a particular field, define its macro to expand
%%% to an empty string, or better, \unskip, like this:
%%%
%%% \newcommand{\showDOI}[1]{\unskip}   % LaTeX syntax
%%%
%%% \def \showDOI #1{\unskip}           % plain TeX syntax
%%%
%%% ====================================================================

\ifx \showCODEN    \undefined \def \showCODEN     #1{\unskip}     \fi
\ifx \showDOI      \undefined \def \showDOI       #1{#1}\fi
\ifx \showISBNx    \undefined \def \showISBNx     #1{\unskip}     \fi
\ifx \showISBNxiii \undefined \def \showISBNxiii  #1{\unskip}     \fi
\ifx \showISSN     \undefined \def \showISSN      #1{\unskip}     \fi
\ifx \showLCCN     \undefined \def \showLCCN      #1{\unskip}     \fi
\ifx \shownote     \undefined \def \shownote      #1{#1}          \fi
\ifx \showarticletitle \undefined \def \showarticletitle #1{#1}   \fi
\ifx \showURL      \undefined \def \showURL       {\relax}        \fi
% The following commands are used for tagged output and should be
% invisible to TeX
\providecommand\bibfield[2]{#2}
\providecommand\bibinfo[2]{#2}
\providecommand\natexlab[1]{#1}
\providecommand\showeprint[2][]{arXiv:#2}

\bibitem[\protect\citeauthoryear{Agarwal, Takatsu, Zaitsev, and
  Joachims}{Agarwal et~al\mbox{.}}{2019a}]%
        {10.1145/3331184.3331202}
\bibfield{author}{\bibinfo{person}{Aman Agarwal}, \bibinfo{person}{Kenta
  Takatsu}, \bibinfo{person}{Ivan Zaitsev}, {and} \bibinfo{person}{Thorsten
  Joachims}.} \bibinfo{year}{2019}\natexlab{a}.
\newblock \showarticletitle{A General Framework for Counterfactual
  Learning-to-Rank}. In \bibinfo{booktitle}{\emph{Proceedings of the 42nd
  International ACM SIGIR Conference on Research and Development in Information
  Retrieval}} (Paris, France) \emph{(\bibinfo{series}{SIGIR'19})}.
  \bibinfo{publisher}{Association for Computing Machinery},
  \bibinfo{address}{New York, NY, USA}, \bibinfo{pages}{5–14}.
\newblock
\showISBNx{9781450361729}
\urldef\tempurl%
\url{https://doi.org/10.1145/3331184.3331202}
\showDOI{\tempurl}


\bibitem[\protect\citeauthoryear{Agarwal, Wang, Li, Bendersky, and
  Najork}{Agarwal et~al\mbox{.}}{2019b}]%
        {10.1145/3308558.3313697}
\bibfield{author}{\bibinfo{person}{Aman Agarwal}, \bibinfo{person}{Xuanhui
  Wang}, \bibinfo{person}{Cheng Li}, \bibinfo{person}{Michael Bendersky}, {and}
  \bibinfo{person}{Marc Najork}.} \bibinfo{year}{2019}\natexlab{b}.
\newblock \showarticletitle{Addressing Trust Bias for Unbiased
  Learning-to-Rank}. In \bibinfo{booktitle}{\emph{The World Wide Web
  Conference}} (San Francisco, CA, USA) \emph{(\bibinfo{series}{WWW '19})}.
  \bibinfo{publisher}{Association for Computing Machinery},
  \bibinfo{address}{New York, NY, USA}, \bibinfo{pages}{4–14}.
\newblock
\showISBNx{9781450366748}
\urldef\tempurl%
\url{https://doi.org/10.1145/3308558.3313697}
\showDOI{\tempurl}


\bibitem[\protect\citeauthoryear{Agarwal, Zaitsev, Wang, Li, Najork, and
  Joachims}{Agarwal et~al\mbox{.}}{2019c}]%
        {10.1145/3289600.3291017}
\bibfield{author}{\bibinfo{person}{Aman Agarwal}, \bibinfo{person}{Ivan
  Zaitsev}, \bibinfo{person}{Xuanhui Wang}, \bibinfo{person}{Cheng Li},
  \bibinfo{person}{Marc Najork}, {and} \bibinfo{person}{Thorsten Joachims}.}
  \bibinfo{year}{2019}\natexlab{c}.
\newblock \showarticletitle{Estimating Position Bias without Intrusive
  Interventions}. In \bibinfo{booktitle}{\emph{Proceedings of the Twelfth ACM
  International Conference on Web Search and Data Mining}} (Melbourne VIC,
  Australia) \emph{(\bibinfo{series}{WSDM '19})}.
  \bibinfo{publisher}{Association for Computing Machinery},
  \bibinfo{address}{New York, NY, USA}, \bibinfo{pages}{474–482}.
\newblock
\showISBNx{9781450359405}
\urldef\tempurl%
\url{https://doi.org/10.1145/3289600.3291017}
\showDOI{\tempurl}


\bibitem[\protect\citeauthoryear{Ai, Bi, Guo, and Croft}{Ai
  et~al\mbox{.}}{2018a}]%
        {10.1145/3209978.3209985}
\bibfield{author}{\bibinfo{person}{Qingyao Ai}, \bibinfo{person}{Keping Bi},
  \bibinfo{person}{Jiafeng Guo}, {and} \bibinfo{person}{W.~Bruce Croft}.}
  \bibinfo{year}{2018}\natexlab{a}.
\newblock \showarticletitle{Learning a Deep Listwise Context Model for Ranking
  Refinement}. In \bibinfo{booktitle}{\emph{The 41st International ACM SIGIR
  Conference on Research \&amp; Development in Information Retrieval}} (Ann
  Arbor, MI, USA) \emph{(\bibinfo{series}{SIGIR '18})}.
  \bibinfo{publisher}{Association for Computing Machinery},
  \bibinfo{address}{New York, NY, USA}, \bibinfo{pages}{135–144}.
\newblock
\showISBNx{9781450356572}
\urldef\tempurl%
\url{https://doi.org/10.1145/3209978.3209985}
\showDOI{\tempurl}


\bibitem[\protect\citeauthoryear{Ai, Bi, Luo, Guo, and Croft}{Ai
  et~al\mbox{.}}{2018b}]%
        {10.1145/3209978.3209986}
\bibfield{author}{\bibinfo{person}{Qingyao Ai}, \bibinfo{person}{Keping Bi},
  \bibinfo{person}{Cheng Luo}, \bibinfo{person}{Jiafeng Guo}, {and}
  \bibinfo{person}{W.~Bruce Croft}.} \bibinfo{year}{2018}\natexlab{b}.
\newblock \showarticletitle{Unbiased Learning to Rank with Unbiased Propensity
  Estimation}. In \bibinfo{booktitle}{\emph{The 41st International ACM SIGIR
  Conference on Research \&amp; Development in Information Retrieval}} (Ann
  Arbor, MI, USA) \emph{(\bibinfo{series}{SIGIR '18})}.
  \bibinfo{publisher}{Association for Computing Machinery},
  \bibinfo{address}{New York, NY, USA}, \bibinfo{pages}{385–394}.
\newblock
\showISBNx{9781450356572}
\urldef\tempurl%
\url{https://doi.org/10.1145/3209978.3209986}
\showDOI{\tempurl}


\bibitem[\protect\citeauthoryear{Ai, Mao, Liu, and Croft}{Ai
  et~al\mbox{.}}{2018c}]%
        {Ai:2018:ULR:3269206.3274274}
\bibfield{author}{\bibinfo{person}{Qingyao Ai}, \bibinfo{person}{Jiaxin Mao},
  \bibinfo{person}{Yiqun Liu}, {and} \bibinfo{person}{W.~Bruce Croft}.}
  \bibinfo{year}{2018}\natexlab{c}.
\newblock \showarticletitle{Unbiased Learning to Rank: Theory and Practice}. In
  \bibinfo{booktitle}{\emph{Proceedings of the 27th ACM International
  Conference on Information and Knowledge Management}} (Torino, Italy)
  \emph{(\bibinfo{series}{CIKM '18})}. \bibinfo{publisher}{ACM},
  \bibinfo{address}{New York, NY, USA}, \bibinfo{pages}{2305--2306}.
\newblock
\showISBNx{978-1-4503-6014-2}
\urldef\tempurl%
\url{https://doi.org/10.1145/3269206.3274274}
\showDOI{\tempurl}


\bibitem[\protect\citeauthoryear{Ai, Wang, Bruch, Golbandi, Bendersky, and
  Najork}{Ai et~al\mbox{.}}{2019}]%
        {10.1145/3341981.3344218}
\bibfield{author}{\bibinfo{person}{Qingyao Ai}, \bibinfo{person}{Xuanhui Wang},
  \bibinfo{person}{Sebastian Bruch}, \bibinfo{person}{Nadav Golbandi},
  \bibinfo{person}{Michael Bendersky}, {and} \bibinfo{person}{Marc Najork}.}
  \bibinfo{year}{2019}\natexlab{}.
\newblock \showarticletitle{Learning Groupwise Multivariate Scoring Functions
  Using Deep Neural Networks}. In \bibinfo{booktitle}{\emph{Proceedings of the
  2019 ACM SIGIR International Conference on Theory of Information Retrieval}}
  (Santa Clara, CA, USA) \emph{(\bibinfo{series}{ICTIR '19})}.
  \bibinfo{publisher}{Association for Computing Machinery},
  \bibinfo{address}{New York, NY, USA}, \bibinfo{pages}{85–92}.
\newblock
\showISBNx{9781450368810}
\urldef\tempurl%
\url{https://doi.org/10.1145/3341981.3344218}
\showDOI{\tempurl}


\bibitem[\protect\citeauthoryear{Ai, Yang, Wang, and Mao}{Ai
  et~al\mbox{.}}{2021}]%
        {10.1145/3439861}
\bibfield{author}{\bibinfo{person}{Qingyao Ai}, \bibinfo{person}{Tao Yang},
  \bibinfo{person}{Huazheng Wang}, {and} \bibinfo{person}{Jiaxin Mao}.}
  \bibinfo{year}{2021}\natexlab{}.
\newblock \showarticletitle{Unbiased Learning to Rank: Online or Offline?}
\newblock \bibinfo{journal}{\emph{ACM Trans. Inf. Syst.}} \bibinfo{volume}{39},
  \bibinfo{number}{2}, Article \bibinfo{articleno}{21} (\bibinfo{date}{Feb.}
  \bibinfo{year}{2021}), \bibinfo{numpages}{29}~pages.
\newblock
\showISSN{1046-8188}
\urldef\tempurl%
\url{https://doi.org/10.1145/3439861}
\showDOI{\tempurl}


\bibitem[\protect\citeauthoryear{Ba, Kiros, and Hinton}{Ba
  et~al\mbox{.}}{2016}]%
        {ba2016layer}
\bibfield{author}{\bibinfo{person}{Jimmy~Lei Ba}, \bibinfo{person}{Jamie~Ryan
  Kiros}, {and} \bibinfo{person}{Geoffrey~E. Hinton}.}
  \bibinfo{year}{2016}\natexlab{}.
\newblock \bibinfo{title}{Layer Normalization}.
\newblock
\newblock
\showeprint[arxiv]{1607.06450}~[stat.ML]


\bibitem[\protect\citeauthoryear{Burges, Shaked, Renshaw, Lazier, Deeds,
  Hamilton, and Hullender}{Burges et~al\mbox{.}}{2005}]%
        {10.1145/1102351.1102363}
\bibfield{author}{\bibinfo{person}{Chris Burges}, \bibinfo{person}{Tal Shaked},
  \bibinfo{person}{Erin Renshaw}, \bibinfo{person}{Ari Lazier},
  \bibinfo{person}{Matt Deeds}, \bibinfo{person}{Nicole Hamilton}, {and}
  \bibinfo{person}{Greg Hullender}.} \bibinfo{year}{2005}\natexlab{}.
\newblock \showarticletitle{Learning to Rank Using Gradient Descent}. In
  \bibinfo{booktitle}{\emph{Proceedings of the 22nd International Conference on
  Machine Learning}} (Bonn, Germany) \emph{(\bibinfo{series}{ICML '05})}.
  \bibinfo{publisher}{Association for Computing Machinery},
  \bibinfo{address}{New York, NY, USA}, \bibinfo{pages}{89–96}.
\newblock
\showISBNx{1595931805}
\urldef\tempurl%
\url{https://doi.org/10.1145/1102351.1102363}
\showDOI{\tempurl}


\bibitem[\protect\citeauthoryear{Burges}{Burges}{2010}]%
        {burges2010from}
\bibfield{author}{\bibinfo{person}{Chris~J.C. Burges}.}
  \bibinfo{year}{2010}\natexlab{}.
\newblock \bibinfo{booktitle}{\emph{From RankNet to LambdaRank to LambdaMART:
  An Overview}}.
\newblock \bibinfo{type}{{T}echnical {R}eport} MSR-TR-2010-82.
\newblock
\urldef\tempurl%
\url{https://www.microsoft.com/en-us/research/publication/from-ranknet-to-lambdarank-to-lambdamart-an-overview/}
\showURL{%
\tempurl}


\bibitem[\protect\citeauthoryear{Cao, Qin, Liu, Tsai, and Li}{Cao
  et~al\mbox{.}}{2007}]%
        {10.1145/1273496.1273513}
\bibfield{author}{\bibinfo{person}{Zhe Cao}, \bibinfo{person}{Tao Qin},
  \bibinfo{person}{Tie-Yan Liu}, \bibinfo{person}{Ming-Feng Tsai}, {and}
  \bibinfo{person}{Hang Li}.} \bibinfo{year}{2007}\natexlab{}.
\newblock \showarticletitle{Learning to Rank: From Pairwise Approach to
  Listwise Approach}. In \bibinfo{booktitle}{\emph{Proceedings of the 24th
  International Conference on Machine Learning}} (Corvalis, Oregon, USA)
  \emph{(\bibinfo{series}{ICML '07})}. \bibinfo{publisher}{Association for
  Computing Machinery}, \bibinfo{address}{New York, NY, USA},
  \bibinfo{pages}{129–136}.
\newblock
\showISBNx{9781595937933}
\urldef\tempurl%
\url{https://doi.org/10.1145/1273496.1273513}
\showDOI{\tempurl}


\bibitem[\protect\citeauthoryear{Chen, Ai, Jayasinghe, and Croft}{Chen
  et~al\mbox{.}}{2019}]%
        {10.1145/3357384.3358131}
\bibfield{author}{\bibinfo{person}{Ruey-Cheng Chen}, \bibinfo{person}{Qingyao
  Ai}, \bibinfo{person}{Gaya Jayasinghe}, {and} \bibinfo{person}{W.~Bruce
  Croft}.} \bibinfo{year}{2019}\natexlab{}.
\newblock \showarticletitle{Correcting for Recency Bias in Job Recommendation}.
  In \bibinfo{booktitle}{\emph{Proceedings of the 28th ACM International
  Conference on Information and Knowledge Management}} (Beijing, China)
  \emph{(\bibinfo{series}{CIKM '19})}. \bibinfo{publisher}{Association for
  Computing Machinery}, \bibinfo{address}{New York, NY, USA},
  \bibinfo{pages}{2185–2188}.
\newblock
\showISBNx{9781450369763}
\urldef\tempurl%
\url{https://doi.org/10.1145/3357384.3358131}
\showDOI{\tempurl}


\bibitem[\protect\citeauthoryear{Hu, Wang, Peng, and Li}{Hu
  et~al\mbox{.}}{2019}]%
        {10.1145/3308558.3313447}
\bibfield{author}{\bibinfo{person}{Ziniu Hu}, \bibinfo{person}{Yang Wang},
  \bibinfo{person}{Qu Peng}, {and} \bibinfo{person}{Hang Li}.}
  \bibinfo{year}{2019}\natexlab{}.
\newblock \showarticletitle{Unbiased LambdaMART: An Unbiased Pairwise
  Learning-to-Rank Algorithm}. In \bibinfo{booktitle}{\emph{The World Wide Web
  Conference}} (San Francisco, CA, USA) \emph{(\bibinfo{series}{WWW '19})}.
  \bibinfo{publisher}{Association for Computing Machinery},
  \bibinfo{address}{New York, NY, USA}, \bibinfo{pages}{2830–2836}.
\newblock
\showISBNx{9781450366748}
\urldef\tempurl%
\url{https://doi.org/10.1145/3308558.3313447}
\showDOI{\tempurl}


\bibitem[\protect\citeauthoryear{Joachims}{Joachims}{2002}]%
        {10.1145/775047.775067}
\bibfield{author}{\bibinfo{person}{Thorsten Joachims}.}
  \bibinfo{year}{2002}\natexlab{}.
\newblock \showarticletitle{Optimizing Search Engines Using Clickthrough Data}.
  In \bibinfo{booktitle}{\emph{Proceedings of the Eighth ACM SIGKDD
  International Conference on Knowledge Discovery and Data Mining}} (Edmonton,
  Alberta, Canada) \emph{(\bibinfo{series}{KDD '02})}.
  \bibinfo{publisher}{Association for Computing Machinery},
  \bibinfo{address}{New York, NY, USA}, \bibinfo{pages}{133–142}.
\newblock
\showISBNx{158113567X}
\urldef\tempurl%
\url{https://doi.org/10.1145/775047.775067}
\showDOI{\tempurl}


\bibitem[\protect\citeauthoryear{Joachims, Granka, Pan, Hembrooke, and
  Gay}{Joachims et~al\mbox{.}}{2005}]%
        {10.1145/1076034.1076063}
\bibfield{author}{\bibinfo{person}{Thorsten Joachims}, \bibinfo{person}{Laura
  Granka}, \bibinfo{person}{Bing Pan}, \bibinfo{person}{Helene Hembrooke},
  {and} \bibinfo{person}{Geri Gay}.} \bibinfo{year}{2005}\natexlab{}.
\newblock \showarticletitle{Accurately Interpreting Clickthrough Data as
  Implicit Feedback}. In \bibinfo{booktitle}{\emph{Proceedings of the 28th
  Annual International ACM SIGIR Conference on Research and Development in
  Information Retrieval}} (Salvador, Brazil) \emph{(\bibinfo{series}{SIGIR
  '05})}. \bibinfo{publisher}{Association for Computing Machinery},
  \bibinfo{address}{New York, NY, USA}, \bibinfo{pages}{154–161}.
\newblock
\showISBNx{1595930345}
\urldef\tempurl%
\url{https://doi.org/10.1145/1076034.1076063}
\showDOI{\tempurl}


\bibitem[\protect\citeauthoryear{Joachims, Granka, Pan, Hembrooke, and
  Gay}{Joachims et~al\mbox{.}}{2017a}]%
        {10.1145/3130332.3130334}
\bibfield{author}{\bibinfo{person}{Thorsten Joachims}, \bibinfo{person}{Laura
  Granka}, \bibinfo{person}{Bing Pan}, \bibinfo{person}{Helene Hembrooke},
  {and} \bibinfo{person}{Geri Gay}.} \bibinfo{year}{2017}\natexlab{a}.
\newblock \showarticletitle{Accurately Interpreting Clickthrough Data as
  Implicit Feedback}.
\newblock \bibinfo{journal}{\emph{SIGIR Forum}} \bibinfo{volume}{51},
  \bibinfo{number}{1} (\bibinfo{date}{Aug.} \bibinfo{year}{2017}),
  \bibinfo{pages}{4–11}.
\newblock
\showISSN{0163-5840}
\urldef\tempurl%
\url{https://doi.org/10.1145/3130332.3130334}
\showDOI{\tempurl}


\bibitem[\protect\citeauthoryear{Joachims, Granka, Pan, Hembrooke, Radlinski,
  and Gay}{Joachims et~al\mbox{.}}{2007}]%
        {10.1145/1229179.1229181}
\bibfield{author}{\bibinfo{person}{Thorsten Joachims}, \bibinfo{person}{Laura
  Granka}, \bibinfo{person}{Bing Pan}, \bibinfo{person}{Helene Hembrooke},
  \bibinfo{person}{Filip Radlinski}, {and} \bibinfo{person}{Geri Gay}.}
  \bibinfo{year}{2007}\natexlab{}.
\newblock \showarticletitle{Evaluating the Accuracy of Implicit Feedback from
  Clicks and Query Reformulations in Web Search}.
\newblock \bibinfo{journal}{\emph{ACM Trans. Inf. Syst.}} \bibinfo{volume}{25},
  \bibinfo{number}{2} (\bibinfo{date}{April} \bibinfo{year}{2007}),
  \bibinfo{pages}{7–es}.
\newblock
\showISSN{1046-8188}
\urldef\tempurl%
\url{https://doi.org/10.1145/1229179.1229181}
\showDOI{\tempurl}


\bibitem[\protect\citeauthoryear{Joachims, Swaminathan, and Schnabel}{Joachims
  et~al\mbox{.}}{2017b}]%
        {10.1145/3018661.3018699}
\bibfield{author}{\bibinfo{person}{Thorsten Joachims}, \bibinfo{person}{Adith
  Swaminathan}, {and} \bibinfo{person}{Tobias Schnabel}.}
  \bibinfo{year}{2017}\natexlab{b}.
\newblock \showarticletitle{Unbiased Learning-to-Rank with Biased Feedback}. In
  \bibinfo{booktitle}{\emph{Proceedings of the Tenth ACM International
  Conference on Web Search and Data Mining}} (Cambridge, United Kingdom)
  \emph{(\bibinfo{series}{WSDM '17})}. \bibinfo{publisher}{Association for
  Computing Machinery}, \bibinfo{address}{New York, NY, USA},
  \bibinfo{pages}{781–789}.
\newblock
\showISBNx{9781450346757}
\urldef\tempurl%
\url{https://doi.org/10.1145/3018661.3018699}
\showDOI{\tempurl}


\bibitem[\protect\citeauthoryear{Li, Burges, and Wu}{Li et~al\mbox{.}}{2007}]%
        {inproceedings}
\bibfield{author}{\bibinfo{person}{Ping Li}, \bibinfo{person}{Christopher
  Burges}, {and} \bibinfo{person}{Qiang Wu}.} \bibinfo{year}{2007}\natexlab{}.
\newblock \showarticletitle{McRank: Learning to Rank Using Multiple
  Classification and Gradient Boosting}.
\newblock \bibinfo{journal}{\emph{Advances in Neural Information Processing
  Systems}}.
\newblock


\bibitem[\protect\citeauthoryear{Liu}{Liu}{2009}]%
        {10.1561/1500000016}
\bibfield{author}{\bibinfo{person}{Tie-Yan Liu}.}
  \bibinfo{year}{2009}\natexlab{}.
\newblock \showarticletitle{Learning to Rank for Information Retrieval}.
\newblock \bibinfo{journal}{\emph{Found. Trends Inf. Retr.}}
  \bibinfo{volume}{3}, \bibinfo{number}{3} (\bibinfo{date}{March}
  \bibinfo{year}{2009}), \bibinfo{pages}{225–331}.
\newblock
\showISSN{1554-0669}
\urldef\tempurl%
\url{https://doi.org/10.1561/1500000016}
\showDOI{\tempurl}


\bibitem[\protect\citeauthoryear{Mao, Luo, Zhang, and Ma}{Mao
  et~al\mbox{.}}{2018}]%
        {10.1145/3209978.3210060}
\bibfield{author}{\bibinfo{person}{Jiaxin Mao}, \bibinfo{person}{Cheng Luo},
  \bibinfo{person}{Min Zhang}, {and} \bibinfo{person}{Shaoping Ma}.}
  \bibinfo{year}{2018}\natexlab{}.
\newblock \showarticletitle{Constructing Click Models for Mobile Search}. In
  \bibinfo{booktitle}{\emph{The 41st International ACM SIGIR Conference on
  Research \&amp; Development in Information Retrieval}} (Ann Arbor, MI, USA)
  \emph{(\bibinfo{series}{SIGIR '18})}. \bibinfo{publisher}{Association for
  Computing Machinery}, \bibinfo{address}{New York, NY, USA},
  \bibinfo{pages}{775–784}.
\newblock
\showISBNx{9781450356572}
\urldef\tempurl%
\url{https://doi.org/10.1145/3209978.3210060}
\showDOI{\tempurl}


\bibitem[\protect\citeauthoryear{O'Brien and Keane}{O'Brien and Keane}{2006}]%
        {O'Brien2006}
\bibfield{author}{\bibinfo{person}{Maeve O'Brien} {and} \bibinfo{person}{Mark
  Keane}.} \bibinfo{year}{2006}\natexlab{}.
\newblock \showarticletitle{Modeling Result-List Searching in the World Wide
  Web: The Role of Relevance Topologies and Trust Bias}.
\newblock  (\bibinfo{date}{01} \bibinfo{year}{2006}).
\newblock


\bibitem[\protect\citeauthoryear{Oosterhuis and de~Rijke}{Oosterhuis and
  de~Rijke}{2018}]%
        {Oosterhuis_2018}
\bibfield{author}{\bibinfo{person}{Harrie Oosterhuis} {and}
  \bibinfo{person}{Maarten de Rijke}.} \bibinfo{year}{2018}\natexlab{}.
\newblock \showarticletitle{Differentiable Unbiased Online Learning to Rank}.
\newblock \bibinfo{journal}{\emph{Proceedings of the 27th ACM International
  Conference on Information and Knowledge Management}} (\bibinfo{date}{Oct}
  \bibinfo{year}{2018}).
\newblock
\showISBNx{9781450360142}
\urldef\tempurl%
\url{https://doi.org/10.1145/3269206.3271686}
\showDOI{\tempurl}


\bibitem[\protect\citeauthoryear{Pang, Xu, Ai, Lan, Cheng, and Wen}{Pang
  et~al\mbox{.}}{2020}]%
        {10.1145/3397271.3401104}
\bibfield{author}{\bibinfo{person}{Liang Pang}, \bibinfo{person}{Jun Xu},
  \bibinfo{person}{Qingyao Ai}, \bibinfo{person}{Yanyan Lan},
  \bibinfo{person}{Xueqi Cheng}, {and} \bibinfo{person}{Jirong Wen}.}
  \bibinfo{year}{2020}\natexlab{}.
\newblock \showarticletitle{SetRank: Learning a Permutation-Invariant Ranking
  Model for Information Retrieval}. In \bibinfo{booktitle}{\emph{Proceedings of
  the 43rd International ACM SIGIR Conference on Research and Development in
  Information Retrieval}} (Virtual Event, China) \emph{(\bibinfo{series}{SIGIR
  '20})}. \bibinfo{publisher}{Association for Computing Machinery},
  \bibinfo{address}{New York, NY, USA}, \bibinfo{pages}{499–508}.
\newblock
\showISBNx{9781450380164}
\urldef\tempurl%
\url{https://doi.org/10.1145/3397271.3401104}
\showDOI{\tempurl}


\bibitem[\protect\citeauthoryear{Pasumarthi, Wang, Bendersky, and
  Najork}{Pasumarthi et~al\mbox{.}}{2019}]%
        {pasumarthi2019selfattentive}
\bibfield{author}{\bibinfo{person}{Rama~Kumar Pasumarthi},
  \bibinfo{person}{Xuanhui Wang}, \bibinfo{person}{Michael Bendersky}, {and}
  \bibinfo{person}{Marc Najork}.} \bibinfo{year}{2019}\natexlab{}.
\newblock \bibinfo{title}{Self-Attentive Document Interaction Networks for
  Permutation Equivariant Ranking}.
\newblock
\newblock
\showeprint[arxiv]{1910.09676}~[cs.IR]


\bibitem[\protect\citeauthoryear{Richardson, Dominowska, and Ragno}{Richardson
  et~al\mbox{.}}{2007}]%
        {10.1145/1242572.1242643}
\bibfield{author}{\bibinfo{person}{Matthew Richardson}, \bibinfo{person}{Ewa
  Dominowska}, {and} \bibinfo{person}{Robert Ragno}.}
  \bibinfo{year}{2007}\natexlab{}.
\newblock \showarticletitle{Predicting Clicks: Estimating the Click-through
  Rate for New Ads}. In \bibinfo{booktitle}{\emph{Proceedings of the 16th
  International Conference on World Wide Web}} (Banff, Alberta, Canada)
  \emph{(\bibinfo{series}{WWW '07})}. \bibinfo{publisher}{Association for
  Computing Machinery}, \bibinfo{address}{New York, NY, USA},
  \bibinfo{pages}{521–530}.
\newblock
\showISBNx{9781595936547}
\urldef\tempurl%
\url{https://doi.org/10.1145/1242572.1242643}
\showDOI{\tempurl}


\bibitem[\protect\citeauthoryear{Robertson and Walker}{Robertson and
  Walker}{1994}]%
        {10.1007/978-1-4471-2099-5_24}
\bibfield{author}{\bibinfo{person}{S.~E. Robertson} {and} \bibinfo{person}{S.
  Walker}.} \bibinfo{year}{1994}\natexlab{}.
\newblock \showarticletitle{Some Simple Effective Approximations to the
  2-Poisson Model for Probabilistic Weighted Retrieval}. In
  \bibinfo{booktitle}{\emph{SIGIR '94}},
  \bibfield{editor}{\bibinfo{person}{Bruce~W. Croft} {and}
  \bibinfo{person}{C.~J. van Rijsbergen}} (Eds.). \bibinfo{publisher}{Springer
  London}, \bibinfo{address}{London}, \bibinfo{pages}{232--241}.
\newblock
\showISBNx{978-1-4471-2099-5}


\bibitem[\protect\citeauthoryear{Schuth, Oosterhuis, Whiteson, and
  de~Rijke}{Schuth et~al\mbox{.}}{2016}]%
        {10.1145/2835776.2835804}
\bibfield{author}{\bibinfo{person}{Anne Schuth}, \bibinfo{person}{Harrie
  Oosterhuis}, \bibinfo{person}{Shimon Whiteson}, {and}
  \bibinfo{person}{Maarten de Rijke}.} \bibinfo{year}{2016}\natexlab{}.
\newblock \showarticletitle{Multileave Gradient Descent for Fast Online
  Learning to Rank}. In \bibinfo{booktitle}{\emph{Proceedings of the Ninth ACM
  International Conference on Web Search and Data Mining}} (San Francisco,
  California, USA) \emph{(\bibinfo{series}{WSDM '16})}.
  \bibinfo{publisher}{Association for Computing Machinery},
  \bibinfo{address}{New York, NY, USA}, \bibinfo{pages}{457–466}.
\newblock
\showISBNx{9781450337168}
\urldef\tempurl%
\url{https://doi.org/10.1145/2835776.2835804}
\showDOI{\tempurl}


\bibitem[\protect\citeauthoryear{Schuth, Sietsma, Whiteson, Lefortier, and
  de~Rijke}{Schuth et~al\mbox{.}}{2014}]%
        {10.1145/2661829.2661952}
\bibfield{author}{\bibinfo{person}{Anne Schuth}, \bibinfo{person}{Floor
  Sietsma}, \bibinfo{person}{Shimon Whiteson}, \bibinfo{person}{Damien
  Lefortier}, {and} \bibinfo{person}{Maarten de Rijke}.}
  \bibinfo{year}{2014}\natexlab{}.
\newblock \showarticletitle{Multileaved Comparisons for Fast Online
  Evaluation}. In \bibinfo{booktitle}{\emph{Proceedings of the 23rd ACM
  International Conference on Conference on Information and Knowledge
  Management}} (Shanghai, China) \emph{(\bibinfo{series}{CIKM '14})}.
  \bibinfo{publisher}{Association for Computing Machinery},
  \bibinfo{address}{New York, NY, USA}, \bibinfo{pages}{71–80}.
\newblock
\showISBNx{9781450325981}
\urldef\tempurl%
\url{https://doi.org/10.1145/2661829.2661952}
\showDOI{\tempurl}


\bibitem[\protect\citeauthoryear{Wang, Liu, Zhang, Ma, Zheng, Qian, and
  Zhang}{Wang et~al\mbox{.}}{2013}]%
        {10.1145/2484028.2484036}
\bibfield{author}{\bibinfo{person}{Chao Wang}, \bibinfo{person}{Yiqun Liu},
  \bibinfo{person}{Min Zhang}, \bibinfo{person}{Shaoping Ma},
  \bibinfo{person}{Meihong Zheng}, \bibinfo{person}{Jing Qian}, {and}
  \bibinfo{person}{Kuo Zhang}.} \bibinfo{year}{2013}\natexlab{}.
\newblock \showarticletitle{Incorporating Vertical Results into Search Click
  Models}. In \bibinfo{booktitle}{\emph{Proceedings of the 36th International
  ACM SIGIR Conference on Research and Development in Information Retrieval}}
  (Dublin, Ireland) \emph{(\bibinfo{series}{SIGIR '13})}.
  \bibinfo{publisher}{Association for Computing Machinery},
  \bibinfo{address}{New York, NY, USA}, \bibinfo{pages}{503–512}.
\newblock
\showISBNx{9781450320344}
\urldef\tempurl%
\url{https://doi.org/10.1145/2484028.2484036}
\showDOI{\tempurl}


\bibitem[\protect\citeauthoryear{Wang, Langley, Kim, McCord-Snook, and
  Wang}{Wang et~al\mbox{.}}{2018b}]%
        {10.1145/3209978.3210045}
\bibfield{author}{\bibinfo{person}{Huazheng Wang}, \bibinfo{person}{Ramsey
  Langley}, \bibinfo{person}{Sonwoo Kim}, \bibinfo{person}{Eric McCord-Snook},
  {and} \bibinfo{person}{Hongning Wang}.} \bibinfo{year}{2018}\natexlab{b}.
\newblock \showarticletitle{Efficient Exploration of Gradient Space for Online
  Learning to Rank}. In \bibinfo{booktitle}{\emph{The 41st International ACM
  SIGIR Conference on Research \&amp; Development in Information Retrieval}}
  (Ann Arbor, MI, USA) \emph{(\bibinfo{series}{SIGIR '18})}.
  \bibinfo{publisher}{Association for Computing Machinery},
  \bibinfo{address}{New York, NY, USA}, \bibinfo{pages}{145–154}.
\newblock
\showISBNx{9781450356572}
\urldef\tempurl%
\url{https://doi.org/10.1145/3209978.3210045}
\showDOI{\tempurl}


\bibitem[\protect\citeauthoryear{Wang, Bendersky, Metzler, and Najork}{Wang
  et~al\mbox{.}}{2016}]%
        {10.1145/2911451.2911537}
\bibfield{author}{\bibinfo{person}{Xuanhui Wang}, \bibinfo{person}{Michael
  Bendersky}, \bibinfo{person}{Donald Metzler}, {and} \bibinfo{person}{Marc
  Najork}.} \bibinfo{year}{2016}\natexlab{}.
\newblock \showarticletitle{Learning to Rank with Selection Bias in Personal
  Search}. In \bibinfo{booktitle}{\emph{Proceedings of the 39th International
  ACM SIGIR Conference on Research and Development in Information Retrieval}}
  (Pisa, Italy) \emph{(\bibinfo{series}{SIGIR '16})}.
  \bibinfo{publisher}{Association for Computing Machinery},
  \bibinfo{address}{New York, NY, USA}, \bibinfo{pages}{115–124}.
\newblock
\showISBNx{9781450340694}
\urldef\tempurl%
\url{https://doi.org/10.1145/2911451.2911537}
\showDOI{\tempurl}


\bibitem[\protect\citeauthoryear{Wang, Golbandi, Bendersky, Metzler, and
  Najork}{Wang et~al\mbox{.}}{2018a}]%
        {10.1145/3159652.3159732}
\bibfield{author}{\bibinfo{person}{Xuanhui Wang}, \bibinfo{person}{Nadav
  Golbandi}, \bibinfo{person}{Michael Bendersky}, \bibinfo{person}{Donald
  Metzler}, {and} \bibinfo{person}{Marc Najork}.}
  \bibinfo{year}{2018}\natexlab{a}.
\newblock \showarticletitle{Position Bias Estimation for Unbiased Learning to
  Rank in Personal Search}. In \bibinfo{booktitle}{\emph{Proceedings of the
  Eleventh ACM International Conference on Web Search and Data Mining}} (Marina
  Del Rey, CA, USA) \emph{(\bibinfo{series}{WSDM '18})}.
  \bibinfo{publisher}{Association for Computing Machinery},
  \bibinfo{address}{New York, NY, USA}, \bibinfo{pages}{610–618}.
\newblock
\showISBNx{9781450355810}
\urldef\tempurl%
\url{https://doi.org/10.1145/3159652.3159732}
\showDOI{\tempurl}


\bibitem[\protect\citeauthoryear{Yue and Joachims}{Yue and Joachims}{2009}]%
        {10.1145/1553374.1553527}
\bibfield{author}{\bibinfo{person}{Yisong Yue} {and} \bibinfo{person}{Thorsten
  Joachims}.} \bibinfo{year}{2009}\natexlab{}.
\newblock \showarticletitle{Interactively Optimizing Information Retrieval
  Systems as a Dueling Bandits Problem}. In
  \bibinfo{booktitle}{\emph{Proceedings of the 26th Annual International
  Conference on Machine Learning}} (Montreal, Quebec, Canada)
  \emph{(\bibinfo{series}{ICML '09})}. \bibinfo{publisher}{Association for
  Computing Machinery}, \bibinfo{address}{New York, NY, USA},
  \bibinfo{pages}{1201–1208}.
\newblock
\showISBNx{9781605585161}
\urldef\tempurl%
\url{https://doi.org/10.1145/1553374.1553527}
\showDOI{\tempurl}


\end{thebibliography}
\end{document}